\documentclass{article}
\begin{document}
\title{ Lagrangian Density for the Vacuum}
\author{Zhang XiangChuan\\The National University of Singapore's Department of Physics'
\\Graduate}
\maketitle
\begin{center}
\begin{minipage}{120mm}
\vskip0.8in
\begin{center}{\bf Abstract}
\end{center}
{\hspace{0.25in}In this paper, the Lagrangian density for the vacuum is mainly discussed,
 meanwhile, the matter field, the modified Lorentz transformation
 relations and the reasons for the invariance of the velocity of light in the vacuum are also discussed.}
\end{minipage}
\end{center}
\vskip1in
\section{The problems of the existing theories$^{[1]}$}
\hspace{0.25in}We know, in special relativity, the priciple of
relativity and the principle of the invariance of the velocity of light in
the vacuum(in this paper,  $c$  represents the velocity of light in the
vacuum in special relativity) are two basic assumptions; Although some
successes have been achieved, some problems still exist:\\ (1) are
there certain relations between the two assumptions? no answer.\\ (2) what is the
meaning of vacuum here? according to
Einstein's meaning, the vacuum was then considered to be empty, 
but actually, existing theories 
and experiments have tell us that
there is matter(for example, fields) in the vacuum, because
light is a kind of matter, and there are interactions among
matters, the interactions can definitely affect the velocity of light, thus we
can logically obtain that the velocity of light in the vacuum is
variable, and the phenomenon in which the actual velocity of light 
exceeds  $c$  can exist in the vacuum.\\ (3) the task of physics is
continously to explore the laws of
the natural world, let people understand and make use of them
better. Even if the assumption of the invariance of the velocity of light
in the vacuum is correct, it is still an assumption, not explained;  
Why is c invariant? behind the assumption, there are definitely physical
reasons, these reasons are still not found, from the theory angle,
this is, of course, a problem of special relativity; if one can
explain it, it is sure that a new physics will emerge.\\
\indent In the unified theories, the Higgs fields play an important role,
but in both the standard model and the supersymmetric theories, the Higgs
field Lagrangian densities are given by hand, the reasons for how to
obtain them are not explained, this is
obviously a problem for the theories from the theory angle.\\
\indent The unified theory has been studied for many years, and
many theories have emerged: the electroweak unified theory, the
standard model, the string theory,
M-theory, etc. the research of these theories is still in progress, at
present, the research of the primary theory has begun. Why can so many theories
emerge? why do different problems still exist in different theoiesy? in
my opinion, one or more of the bases of these theories is(are) problematic, for example, 
the Yang-Mill's gauge theory is not
complete. one of the purposes foe me to write this paper is to provide
a new angle(the angle of the level of disorder) for the primary
theory. Besides this paper, I will
discuss some problems related to the unified theory from the angle 
of the level of disorder in my follow-up papers. I hope that I will be
able to construct an acceptable unified theory of the four basic interactions.\\
\indent Now let us discuss how to solve the above-mentioned problems.
\section{Matter Field$^{[2]}$}
\hspace{0.25in}According to the elementary particle theory, in terms of spin, all particles in
the universe can be devided into two categories: one is the boson which
the spin is 0, 1 , 2 , 3 , $\cdots$, the other is the fermion which
the spin is ${1}\over{2}$ ,${3}\over{2}$ ,${5}\over{2}$ ,$\cdots$
According to the quantum field theory, each kind of 
particle corresponds to one kind of field, and is the quantum excited by the field; 
the fermi field  and the bose field can interact each other, this shows that the fermi 
field and the bose field have some common properties; While all matter in the universe is 
composed by fermions and/or bosons, thereby, the universe can be considered as one kind of
field---the matter field which is more elementary than the fermi field
and the bose field. Thus, any particle can be regarded as a matter
field, and is a form of expression of the matter
field; Any matter system can be treated as a matter field, and is a form of expression of 
the matter field. According to the quantum field 
theory, there are four kinds of elementary interactions: the strong 
interaction, the weak interaction, the electromagnetic interaction and the gravitational
interaction. For the strong interaction and the weak 
interaction, the interaction range is very short, for the electromagnetic interaction
and the gravitational interaction, the interaction range can be from
zero to infinity. Hence, I think colour charges and weak charges are not the 
essential properties of the quanta of the matter field, meanwhile
considered that the matter field itself is the source of the
gravitational interaction. thereby  only elertric charges are the essential 
properties of the quanta of the matter field( here, I also mean that
the level of the gravitational interaction and the level of
the electromagnetic interaction are the same, they are different from
that of the weak interaction and the strong interaction, the former is
more elementary), this means that there are only two
kinds of the quanta of the matter field, they are $M_{p}$ with the
positive electric charge and $M_{n}$ with the negative electric charge.\\   
\section{The modified Lorentz Transformation Relations$^{[3]}$}
\hspace{0.25in}All matter in the universe is in a certain space and time, space and time 
mean the space and time of matter, this implies that matter and space,
time depend on and affect each other. Hence, when we 
investigate the physics laws of matter we must consider matter, space and time as a whole, 
only thus, the system considered is a complete system. Hence, when we
set up a reference system, in order to meet this requirement, 
we must set up an abstract four dimensional reference system,
and regard the time axis and the space axes as the isotopic axes, meanwhile, we stipulate 
that the time coordinate is imaginary(for the time axis is virtual), i. e. ikt, where k is 
a constant which is greater than zero, i is an imaginary symbol, t represents the real time,
thus, the axes of this four dimensional reference system are ikt , x , y , z respectively. 
Because the dimension of kt is length, the dimension of $k$ is the velosity.\\
\indent According to the principle of relativity, all inertial reference systems are 
equivalent.{\bf Now that all inertial reference systems are equivalent, and the system
composed by the four dimensional space and matter is complete, so in
the four dimensional space, the four dimensional shape of an object
must be the same in all inertial reference systems}, this means that the distance between any two points in the four dimensional space 
is the same in all inertial reference systems, this is the invariability of the four 
dimensional interval. Thereby, from the mathematical angle, the coordinate
transformation between any two inertial reference systems is a rotation or the combination 
of a rotation and a translation in the four dimensional space, but the translation implies 
that the origins of the two inertial reference systems don't coincide at the beginning,
no other meaning, so, only the rotation needs to be considered. The transformation for a 
four dimensional space rotation is a four dimensional orthogonal transformation.\\
\indent Considering the two inertial reference systems S , S', S moves at a velocity 
$\vec{V}$ relative to S' , for an arbitrary point P(ikt, x, y, z) in S, it is 
corresponding to the point P'(ikt', x', y', z') in S' , thus we
obtain:\\
\begin{eqnarray}  
\left(\begin{array}{c}
ikt'\\x'\\y'\\z'\end{array}\right)   
 & =\left(\begin{array}{llll}
a_{11} & a_{12} & a_{13} & a_{14}\\
a_{21} & a_{22} & a_{23} & a_{24}\\
a_{31} & a_{32} & a_{33} & a_{34}\\
a_{41} & a_{42} & a_{43} & a_{44} 
\end{array}\right) & 
\left(\begin{array}{c} 
ikt\\x\\y\\z\end{array}\right)\end{eqnarray}
\label{e1}
where the transformation matrix $T_{matrix}$  is an orthogonal matrix.\\
Now considering a special case which the axes of S are parallel to the corresponding axes 
of S' , and the corresponding axes of S and S'  have the same directions, when $t=t'=0$, 
the origins of the two coordinate systems coincide, the direction of $\vec{V}$ is the same 
as  the positive direction of x axis of S , thus, the transformation matrix becomes the 
following form:\\
\[\left(\begin{array}{llrr}
a_{11} & a_{12} & 0 & 0\\
a_{21} & a_{22} & 0 & 0\\
0 & 0 & a_{33} & 0\\
0 & 0 & 0 & a_{44}
\end{array}\right)\] \\ 
In terms of the properties of the orthogonal transformation, we obtain the following 
equations:\\
\begin{eqnarray}
a_{33}=a_{44} &  =1 \\ 
a_{11}^2+a_{12}^2 &  =1 \\
a_{11}a_{21}+a_{12}a_{22} &  =0\\
a_{11}^2+a_{21}^2 &  =1
\end{eqnarray}\label{e2,3,4,5}
from this, we obtain:\\
\begin{eqnarray}
a_{11}^2=a_{22}^2\\
a_{12}^2=a_{21}^2
\end{eqnarray}
\label{e6.7}
Considering the origin of S, in S, it menas $x=0$ which is corresponding to $x'=Vt'$ in S' . 
According to equation (1) and the transformation matrix a of the special case, we obtain 
the following equations:\\ 
$ikt'=a_{11}ikt+a_{12}x$\\
$x'=a_{21}ikt+a_{22}x$\\
Hence, we obtain:\\ 
\begin{equation}
a_{11}^2=a_{21}^2{k^{2}\over V^{2}}
\end{equation}
\label{e8}\\ 
According to equations (2), (3), (3), (4), (5), (6), (7), (8), we
obtain:\\
\begin{eqnarray} 
a_{12}^2&={-V^{2}\over {k^{2}-V^{2}}}\\
a_{11}^{2}&={k^{2}\over {k^{2}-V^{2}}} 
\end{eqnarray} 
\label{e9,10}  
Because the positive directions of the axes of the two reference systems are the same,
$a_{11}>0$ , $a_{22}>0$. thereby we obtain:\\
$a_{11}=a_{22}={1\over\sqrt{1-{V^{2}\over k^{2}}}}$ , $k>|\vec{V}|$\\ 
The increment of t can result in the increment of x', and 
$a_{11}a_{21}+a_{12}a_{22}=0$\\
so, $a_{12}=-a_{21}={{iV\over k}\over\sqrt{1-{V^{2}\over k^{2}}}}$\\
Thus, we obtain the transformation relations of the coordinates of the two inertial reference 
systems:\\
\begin{eqnarray}
ikt'={ikt\over\sqrt{1-{V^{2}\over k^{2}}}}+
{{ixV\over k}\over\sqrt{1-{V^{2}\over k^{2}}}}\\ 
x'={-iV\over k}{ikt\over\sqrt{1-{V^{2}\over k^{2}}}}+{x\over\sqrt{1-{V^{2}\over k^{2}}}}\\ 
y'=y\\
z'=z
\end{eqnarray}\label{11,12,13,14}
This is the modified Lorentz transformation relations. Because $\vec {V}$ is 
arbitrary, and $k>V$, k is the maximum realizable velocity in the universe.\\
\indent Please pay attention to the following two points:\\ 
(1) to obtain the modified Lorentz relations, we use the
principle of relativity only, do not use the principle of the
invariance of the velocity of light in the vacuum; this means no certain
relations between the two principles. This also implies that
the principle of the invariance of the velocity of light in the vacuum
is not necessary in special relativity.\\
(2) $k$ is not less than $c$.  
\section{The Reasons Why the Velocity of
Light in the Vacuum Is Constant and  Equal in All Systems Moving with Constant Velocities}  
\hspace{0.25in} In  special relativity, that the velocity of light in the vacuum is 
constant and equal in all systems moving with constant velocities is taken as a basic 
assumption, till now, this assumption has still not been explained, from
the theory angle, this is a problem of special relativity. Now, let us  use the
matter field to explain the assumption.\\ \indent  According
to the disscusions in section two, the quanta of the matter field are
electrically charged, when external electric
charges present, the quanta will be polarized and form a certain
distribution; When the external electric charges change, the
distribution will also change  with the 
change of the external electric charges; When the  external electric
charges oscilate, the change of
the distribution will transmit in a form of wave in space, the wave
 is regarded as the electromagetic wave here. Now let us explain the reasons why
the transmitting velocity of the electromagnetic wave(the polarization wave) in the vacuum 
is constant and equal in all systems moving with constant velocities.\\
\setlength{\unitlength}{0.1in}
\begin{picture}(60,23)(0,0)
\put(0,0){\vector(1,0){50}}
\put(0,0){\vector(0,1){22}}
\put(0,11){\circle*{2}}
\put(2.5,12.1){\makebox(2,2)[l]{B}}
\put(25,2){\circle{4.1}}
\put(35,2){\circle{4.1}}
\put(15,4){\line(1,0){30}}
\put(45,4){\line(0,1){11}}
\put(15,15){\line(1,0){30}}
\put(15,4){\line(0,1){11}}
\put(46,11){\vector(1,0){3}} 
\put(2.5,8.5){\makebox(3,3)[l]{sound source}}
\put(51,0){\makebox(2,2)[l]{X}}
\put(1,21){\makebox(2,2)[l]{Y}}
\put(31,9){\makebox(2,2)[l]{\circle*{0.5}}}
\put(33,9.9){\makebox(2,2)[l]{A}}
\put(27,14.1){\makebox(2,2)[l]{\circle*{0.5}}}
\put(29,15){\makebox(2,2)[l]{C}}
\put(50.5,11.2){\makebox(2,2)[l]{$\vec{V}$}}
\put(23.5,-5.5){\makebox(2,4)[l]{(Fig.1)}}     
\end{picture}
\\
\\
\\
\\
\indent See Fig.1, there is an enough long, enough wide and enough high carriage which has
an open rear side(the left side of the carriage in Fig.1) and the other
closed sides and moves at a velocity $\vec{V}$
along the positive direction of X axis, a sound source is at point B and continuously give 
out sound waves, the transmitting velocity of the sound waves is $\vec{V_{s}}$ in air. 
when the sound waves arrives at observers A , C respectively, because 
observer C is on the top of the carriage, observer C's velocity relative to the
air outside the carriage is also $\vec{V}$, hence, the sound velocity determined by 
ddobserverC is $\vec{V_{s}}-\vec{V}$(here ignoring the effects of  relativity); 
while observer A is in the carriage, observer A is at rest relative to the air inside the 
carriage, thus, when the sound waves transmit into the air inside the carriage, the velocity 
of the sound waves is still $\vec{V_{s}}$ in the air inside the carriage. Thereby, the 
velocity of the sound waves determined by observer A is still $\vec{V_{s}}$, though observer 
A is moving with the carriage, so long as the relation of $|\vec{V_{s}}|>|\vec{V}|$ is 
satisfied.\\
\indent In terms of the same principle, we can explain the reasons why the velocity of 
light in the vacuum is constant and equal in all systems moving with constant velocities. 
Because all matter systems in the universe are matter fields, and are 
the forms of expression of the matter field; Meanwhile, because there are interactions 
between any two matter systems in the universe, there is a matter field surrounding any matter
system, therefore, there is also a matter field in the vacuum. When the matter system 
moves, the matter field surrounding the matter system also moves with the matter system.\\
\indent See Fig.2, supposed that in the vacuum there is a source of light at point A and an  
observer(equivalent to a matter system) at point B, observer $B$ wants
to measure the velocity of light given out by the source of light at
point A. It is obvious that no matter how the relative    
motion between the source of light and observer B is(but must ensure that the light of the    
source A can transmit into the matter field surrounding observer B),
because the matter field surrounding the source of light at point A and the
matter field in the vacuum mix together as well as the matter            
field in the vacuum and the matter field surrounding observer B also mix together, the light 
waves can transmit into the matter field in the vacuum from the matter field surrounding
the source of light at point A, then can transmit into the matter
field surrounding observer B from the matter field in the vacuum; And the     
velocity of light determined by observer B is always the velocity of light in the matter      
field surrounding observer B itself, it is impossible for observer B to
directly measure the velocity of light in the vacuum, it is only
possible for observer B to directly measure the velocity of light in
the matter field surrounding observer B itself, so is it for the other
observers, hence, the velocity of light determined by any  observer is the    
same(supposed the matter fields surrounding all observers are the same), and is
always equal to the velocity of light in the vacuum(supposed the
matter fields surrounding all observers are the same as the matter field in the
vacuum),  
so long as the absolute value of the velocity of light in the vacuum is greater than that of  
the velocity of any observer. Thus we have explained the reasons why
the velocity of light in the vacuum is constant and equal in all
systems moving with constant velocities.In 1887, in order 
to prove that if there was the ether in the vacuum, Michelson-Morley did the famous 
Michelson-Morley experiment, from the matter field's point of view, the experimental 
results are natural. Now iet us discuss the influence on the velocity of
light in the vacuum given by the density of the quanta of the matter
field in the vacuum. because the density can affect the velocity of
light, and the densities surrounding different observers are
different, the velocities of light determined by different    
observers are different, normally, these differences are very small(of
course, if the differences of the densities are big, the differences of the
velocities are big). Set the density to which $c$
corresponds as $DN_{c}$, because it is impossible for the vacuum to be
empty, when the density is not equal to $DN{c}$, the actual velocity
of light in the vacuum is not equal to $c$($>c$ and $<c$), thus
we can obtain that the phenomenon in which the actual velocity of
light exceeds $c$ can exist in the vacuum.\\
\setlength{\unitlength}{0.1in}                                                                
\begin{picture}(60,23)                                                                        
\put(51,0){\makebox(2,2)[l]{X}}
\put(1,21){\makebox(2,2)[l]{Y}}
\put(0,0){\vector(1,0){50}}
\put(0,0){\vector(0,1){22}}
\put(15,10){\circle*{2}}
\put(35,16){\circle{5}}
\put(14.5,12){\makebox(2,2)[l]{A}}
\put(34,19.5){\makebox(2,2)[l]{B}}
\put(11.5,6.5){\makebox(2,2)[l]{Light Source}}
\put(32,11.5){\makebox(2,2)[l]{Observer}}
\put(19,-5.5){\makebox(2,4)[l]{(Fig. 2)}}
\end{picture}
\\
\\
\\
\section{The Lagrangian Density for the Vacuum$^{[4]}$}
\hspace{0.25in}According to the discussions in section two and existing
experimental results, we know that there are the free
quanta of the matter field and the neutral and electrically charged
particles which are composed by the    
quanta of the matter field in the vacuum, these particles are in a dynamic equilibrium state, 
i. e. the vacuum state. the vacuun state is also one state of the  matter system. The state of
a matter system is the distribution of the number density $\rho$(here the
density for space density only) of the quanta of the matter field of
the matter system in the space-time. Different state of the matter system corresponds to 
the different distribution. In order to investigate the change laws of
the state of the matter system, we normally find a few state functions
of the matter system, these state functions describe the  
properties of the state of the matter system from different sides, the density of the level of
disorder(DOLOD for short) is such a state function of the matter system. DOLOD means the grade
which the quanta of the matter field of the matter system in unit
space volume are in chaotic state 
in the space-time, it is a scalar in the space-time. The greater DOLOD, the higher the grade, 
vice versa. There are two kinds of factors in a matter system, one of them makes DOLOD increase, the other makes DOLOD decrease, the natural nature of the matter system always makes DOLOD of itself increase, thereby,
when the matter system is in a dynamic equilibrium state, DOLOD has and reaches a maximum\\
\indent Now we discuss how to quantitatively express DOLOD. Now that the   
state of the matter system is the distrbution of $\rho$, and DOLOD is a state function of the 
matter system, then there is definitely a direct relation between DOLOD and the 
distribution. Meanwhile, from the mathamatical angle, if $\rho$ were 
negative(for example, the number density with positive charges or negative charges), 
DOLOD would be still the same (compared to DOLOD with $+\rho$ which has 
the same absolute value as $-\rho$), so DOLOD must be the function of
${\rho}^2$. 
When $\rho=0$, DOLOD must be zero, because $\rho=0$ implies 
that there is no matter in the space-time. When
$\rho\neq0$(consider number only), 
DOLOD must be a positive value. Thus, DOLOD 
caused by $\rho$ can be expressed as the following form:\\
$W_{1}=W_{1}(\rho^{2})$\\
$W_{1}(0)=0$ \& $\frac{\partial W_{1}(\rho^{2})}{\partial\rho^{2}}|_{0<\rho<\rho_{m}}>0$,     
 where, $\rho_{m}$ is the first extremum point of $W_{1}$, and $\rho_{m}>0$\\                
When $\rho$ has a distribution in the space-time, there is
a certain relation among the $\rho$s at all points in the matter
system, the certain relation can also have contribution to DOLOD, the
contribution is determined by the four dimensional gradient of
$\rho$. Because for any point, no  matter how the direction of the four          
dimensional gradient is, DOLOD is the same, DOLOD must be the function
of the square of the four dimensional gradient; At the same time, 
the greater the four dimensional gradient, the smaller
DOLOD, vice versa. So DOLOD caused by the four dimensional gradient can be expressed as:\\   
$W_{2}=W_{2}((\nabla_{4}\rho )^{2})$,                                                         
where, $\nabla_{4}$ is the symbol of the four dimensional gradient, i. e.\\                  
$\nabla_{4u}=(\frac{\partial}{\partial(ikt)}, {\bf\nabla}),  u=1, 2, 3, 4$\\                 
$(\nabla_{4}\rho )^{2}=(\nabla_{4}\rho )\cdot(\nabla_{4}\rho )$\\                            
This is the case that the matter system is  at rest. When the matter system is in motion,
DOLOD                                                  
of the matter system will decrease, because the motion along a direction reduces the grade of 
chaos of the matter system. DOLOD caused by the motion of the matter system will be given from
the invariability of DOLOD. DOLOD caused by the irregular motion of the quanta of the matter  
field can be involved in DOLOD caused by $\rho$,
because the irregular motion can affect the distribution of $\rho$. 
For a certain matter system, under the same external conditions, the relation
between the distribution of $\rho$ and the irregular motion
one-to-one. In fact, the level of this irregular motion is determined by the  
universe temperature. When a matter system is in a dynamic equilibrium state, the irregular   
motion has a certain level, the physical quantity for expressing the
level is the universe temperature. the higher the universe temperature, the higher the level 
, vice versa; The relation between them is  one-to-one; Thereby, the   
relation between the distribution of $\rho$ and the universe          
temperature is also one-to-one. Except the above-mentioned factors, no other factor can have  
contribution to DOLOD. Thus, when a matter system is at rest, its DOLOD can be expressed as   
the following form:\\                                                                        
$WX=W_{1}(\rho^{2})+W_{2}((\nabla_{4}\rho )^{2})$ \\  
\indent Now we discuss the detailed expressions of $W_{2}((\nabla_{4}\rho)^{2})$ and
$W_{1}(\rho^{2})$. Expanding $W_{2}((\nabla_{4}\rho)^{2})$ into a power series:\\
$W_{2}((\nabla_{4}\rho)^{2})=W_{2c}+k_{22}(\nabla_{4}\rho)^{2}+\cdots$\\
where, $W_{2c}$ , $k_{22}$ are constants.\\
Because $W_{2}((\nabla_{4}\rho )^{2})$ is DOLOD for reflecting the non-uniformity of the
distribution of the number density of the quanta of the matter field inside the matter system,
and the non-uniformity can completely be expressed out by $(\nabla_{4}\rho )^{2}$,            
so we obtain:\\
$W_{2}{(\nabla_{4}\rho )^{2}}=W_{2c}+k_{22}(\nabla_{4}\rho )^{2}$, $k_{22}<0$\\
\indent Because $W_{1}(\rho^{2})$ is DOLOD for reflecting the deviation of the number density 
of the quanta of the matter field relative to the zero point, it is related to the nature of  
the space-time. Expanding $W_{1}(\rho^{2})$ into a power series:\\
$W_{1}(\rho^{2})=W_{1c}+k_{0}\rho^{2}+k_{1}\rho^{4}+k_{2}\rho^{6}+\cdots$\\
where, $W_{1c}$, $k_{0}$, $k_{1}$, $k_{2}$ are constants. thereby,\\
$WX=W_{c}+k_{22}(\nabla_{4}\rho )^{2}+k_{0}\rho^{2}+k_{1}\rho^{4}+k_{2}\rho^{6}+\cdots$\\
where, $W_{c}=W_{1c}+W_{2c}$\\                  
\indent Now considering the case. For the vacuum, there are not only the free quanta of the
matter field but also the neutral and charged particles composed by the quanta of the matter  
field, so, there are four kinds of number densities:\\
$\rho_{s0}$, $\rho_{sq}$, $\rho_{b0}$, $\rho_{bq}$\\
their meanings are as follows respectively:\\         
$\rho_{s0}$: The number density of the free quanta of the matter field
at an arbitrary  point;\\
$\rho_{sq}$: The number density of free quanta of the net charge part of the matter field 
at an arbitrary point;\\                   
$\rho_{b0}$: The number density of the particles which are composed by the quanta of the 
matter field at an arbitrary point(maybe have many kinds of particles, take one as a  
representative here);\\
$\rho_{bq}$: The number density of particles of the net charge part which are composed by the 
quanta of the matter field at an arbitrary point(maybe have many kinds of particles, take one 
as a representative here).\\                              
Hence, $W_{1}$ should be the function of $\rho_{s0}^{2}$, $\rho_{sq}^{2}$,
$\rho_{b0}^{2}$, $\rho_{bq}^{2}$, i. e.:\\              
$W_{1}=W_{1}(\rho_{s0}^{2}, \rho_{sq}^{2}, \rho_{b0}^{2}, \rho_{bq}^{2})$\\
$W_{2}$ should be the function of $(\nabla_{4}\rho_{s0} )^{2}$, $(\nabla_{4}\rho_{sq})^{2}$,
$(\nabla_{4}\rho_{b0} )^{2}$, $(\nabla_{4}\rho_{bq} )^{2}$, i. e.:\\
$W_{2}=W_{2}((\nabla_{4}\rho_{s0} )^{2}, (\nabla_{4}\rho_{sq})^{2}, 
(\nabla_{4}\rho_{b0} )^{2}, (\nabla_{4}\rho_{bq} )^{2})$\\
thus, according to the same principle and method as the
above-mentioned ones, we can obtain:
\\
$W_{1}=W_{1cv}+\lambda_{1}\rho_{s0}^{2}+\lambda_{2}\rho_{sq}^{2}+\lambda_{3}\rho_{b0}^{2}
+\lambda_{4}\rho_{bq}^{2}+\lambda_{5}\rho_{s0}^{4}+\cdots+\lambda_{14}\rho_{bq}^{4}+\cdots$\\
where, $W_{1cv}$, $\lambda_{1}$, $\lambda_{2}$, $\cdots$, $\lambda_{14}$ are constants.\\
$W_{2}=W_{2cv}+h_{1}(\nabla_{4}\rho_{s0})^{2}+h_{2}(\nabla_{4}\rho_{sq})^{2}+
h_{3}(\nabla_{4}\rho_{b0} )^{2}+h_{4}(\nabla_{4}\rho_{bq})^{2}$\\
where, $W_{2cv}$, $h_{1}$, $h_{2}$, $h_{3}$, $h_{4}$ are constants.\\
thereby,\\
\begin{equation}
WX=W_{cv}+W_{1y}+W_{2y}\end{equation}\label{e15}
where, $W_{cv}=W_{1cv}+W_{2cv}$ \\
$W_{1y}=\lambda_{1}\rho_{s0}^{2}+\lambda_{2}\rho_{sq}^{2}+\lambda_{3}\rho_{b0}^{2}
+\lambda_{4}\rho_{bq}^{2}+\lambda_{5}\rho_{s0}^{4}+\cdots+\lambda_{14}\rho_{bq}^{4}+\cdots$\\
$W_{2y}=h_{1}(\nabla_{4}\rho_{s0})^{2}+h_{2}(\nabla_{4}\rho_{sq})^{2}+
h_{3}(\nabla_{4}\rho_{b0} )^{2}+h_{4}(\nabla_{4}\rho_{bq})^{2}$\\
\indent Because $\rho_{s0}$, $\rho_{sq}$, $\rho_{b0}$, $\rho_{bq}$ are the different
ingredients of the vacuun, for the vacuum, they are isotopic. Thus, they can be regarded as
different components of a isovector in the isospace, i. e. :\\
\begin{eqnarray*}
\rho_{v}&=\left(\begin{array}{l}
f_{s0}\rho_{s0}\\f_{sq}\rho_{sq}\\f_{b0}\rho_{b0}\\f_{bq}\rho_{bq}\end{array}\right)
\end{eqnarray*}\\
where, $f_{s0}$, $f_{sq}$, $f_{b0}$, $f_{bq}$ are constants. But in the space-time, $\rho_{v}$
is a scalar. Thus, Thus, we naturally hope that we can express DOLOD
for the vacuum as the  following form:\\                                                     
\begin{equation}WX_{v}=W_{zcv}+\beta _{0}|\nabla_{4}\rho _{v}|^{2}
+\beta_{1}|\rho_{v}|^{2}+\beta_{2}|\rho_{v}|^{4}+\beta_{3}|\rho_{v}|^{6}+\cdots
\end{equation}\label{e16}\\
where, $W_{zcv}$, $\beta_{0}$, $\beta_{1}$, $\beta_{2}$, $\beta_{3}$ are constants,\\
$|\nabla_{4}\rho _{v}|^{2}=(\nabla_{4}\rho _{v})^{\dagger}(\nabla_{4}\rho_{v})$,\\
$|\rho_{v}|^{2}={\rho_{v}}^{\dagger}\rho_{v}$\\
From the mathematical angle, if we expand  $WX_{v}$  into the form of the right side of
equation (15), the total items of the right side of equation (16) are equal to that of
equation (15), hence, so long as the coefficients of the corresponding items of the
two equations are equal, equation (16) can be realized, while this is completely possible.
Here, equation (8) is selected to be the form of expression of DOLOD for the vacuum, because
equation (16) can expresses the entirety and harmonicity of the vacuum better.\\
\indent According to the theory of the solid state physics, the third order derivative and
the above of the potential of atoms are non-harmonic items, they are the reasons for the
expansion caused by heat and contraction caused by cold, if no these derivatives, then no
phenomenon of expansion caused by heat and contraction caused by cold. Here, $W_{1}$ is
equivalent to the potential, $\rho^{2}$ is equivalent to the distance between two atoms. Under
a certain universe temperature, the quanta of the matter field in the universe are in a 
certain
dynamic equilibrium state and have a certain average of the number density at any point. When 
the universe temperature changes, the instantaneous number density of the quanta of the matter
field will change, if the average of the number density is different from that of before the
change of the universe temperature, then the third order derivative and the above of $W_{1}$
are not zero; vice versa. Of course, the change of the universe temperature must be within a
certain range, out of this range, the average of the number density for the dynamic 
equilibrium state will change. In actual situations, what we investigate is a part
of the universe, not the 
whole universe, so we can encounter non-harmonic problems, if what we investigate is the whole
universe, or even if a part of the universe, but we consider all of the relevant factors, we  
will also not encounter non-harmonic problems. Maybe this is one of the meanings that the
universe is harmonic. The harmonic point of view will be sticked to in this paper. Thus, DOLOD
for the vacuum can be expressed as the following form:\\
\begin{equation}WX_{v}=W_{zcv}+
\beta_{0}|\nabla_{4}\rho_{v}|^{2}
+\beta_{1}|\rho_{v}|^{2}+\beta_{2}|\rho_{v}|^{4}\end{equation}\label{e17}\\
\indent Now we discuss the constants of equation (17). For equation (17), in the space-time,
$\rho_{v}$ is a scalar, so, according to the preceding discussion in this section,
$\beta_{0}<0$; when the vacuum matter system is in dynamic equilibrium state, its DOLOD has
and reaches a maximum, so $\beta_{1}>0$,  $\beta_{2}<0$. For the constants in $\rho_{v}$,
consider the expanding form of equation (17) which is the same as the right form of equation
(15), the following conditions must be satified:\\
$f_{s0}>0$,  $f_{b0}>0$ \\
Since the greater the square of the four dimensional gradient of the number density of the net
charge part, the greater DOLOD, vice versa; the greater the square of the number density of
the net charge part, the smaller DOLOD, vice versa, we can also obtain the following 
relations:\\
$f_{sq}^{\dagger}f_{sq}<0$,  $f_{bq}^{\dagger}f_{bq}<0$\\
so $f_{sq}$, $f_{bq}$ must be operators and commutate with $\nabla_{4}$. doing the
replacement:\\
$f_{sq}$ $\rightarrow$ $f_{sq}\stackrel{\wedge}{C_{N}}$,
$f_{bq}$ $\rightarrow$ $f_{bq}\stackrel{\wedge}{C_{N}}$, $f_{sq}>0$, $f_{bq}>0$\\
where, $\stackrel{\wedge}{C_{N}}$ is an operator and commutates with $\nabla_{4}$, and
satisfies the condition:\\
$\stackrel{\wedge}{C_{N}}^{\dagger}\stackrel{\wedge}{C_{N}}=-\stackrel{\wedge}{I}$,
where, $\stackrel{\wedge}{I}$ is the identity operator.\\
This operator is a conversion operator between the number density of a net charge part and
the same neutral number density, called the charge conversion operator. According to
the meaning of the charge conversion operator, it must also satisfy
the following condition:\\
$\stackrel{\wedge}{C_{N}}\stackrel{\wedge}{C_{N}}=\stackrel{\wedge}{I}$,
$\stackrel{\wedge}{C_{N}}^{-1}=\stackrel{\wedge}{C_{N}}$\\
Thus we obtain:\\
\begin{eqnarray*}
\rho_{v}&=\left(\begin{array}{l}
f_{s0}\rho_{s0}\\f_{sq}\stackrel{\wedge}{C_{N}}\rho_{sq}\\f_{b0}\rho_{b0}\\
f_{bq}\stackrel{\wedge}{C_{N}}\rho_{bq}\end{array}\right)
\end{eqnarray*}\\
\indent In terms of the discussions of section three, we can obtain that the length of a 
moving         
object will contract. Supposing the rest length of the object is L, when the object moves
at a velocity $\vec{V}$ along the direction in which the length of the object is measured,
the length of the object in the direction of $\vec{V}$ becomes: 
$L\sqrt{1-{V^{2}\over k^{2}}}$,
so for the vacuum, the effect of length contraction will lead to that the four kinds of number
densities in $\rho_{v}$ will increase to
$1\over\sqrt{1-{V^{2}\over k^{2}}}$ as much as that of at rest
Because all inertial systems are equivalent, and DOLOD is an invariable, we must modify the
expression of DOLOD to meet the requirement. According to equation (17), if $\rho _{v}$ is
a rest physical quantity $\rho _{vr}$, $WX_{v}$ is an invariable. Thus, when the matter
system moves at a velocity $\vec{V}$, the invariable expression of DOLOD is as follows:\\
\begin{equation}WX_{v}=W_{zcv}+\beta_{0}(1-{V^{2}\over {k^{2}}})|\nabla_{4}\rho _{v}|^{2}
+\beta_{1}(1-{V^{2}\over {k^{2}}})|\rho _{v}|^{2}
+\beta_{2}(1-{V^{2}\over
  {k^{2}}})^{2}|\rho_{v}|^{4}\end{equation}\label{e18}\\
According to the meanings of Lagrangian density and DOLOD, from the
angle of the physics essence, Lagrangian
density and DOLOD are equivalent, so the Lagrangian density for the
vacuum is:\\
\begin{equation}L_{v}=W_{zcv}+\beta_{0}(1-{V^{2}\over {k^{2}}})|\nabla_{4}\rho _{v}|^{2}
+\beta_{1}(1-{V^{2}\over {k^{2}}})|\rho _{v}|^{2}
+\beta_{2}(1-{V^{2}\over
  {k^{2}}})^{2}|\rho_{v}|^{4}\end{equation}\label{e19}\\ 
Please pay attention to that this expression is correct for the vacuum
only.
If the matter system is at rest, the Lagrangian density becomes:\\
$L_{v}=W_{zcv}+\beta_{0}|\nabla_{4}\rho_{vr}|^{2}
+\beta_{1}|\rho_{vr}|^{2}+\beta_{2}|\rho_{vr}|^{4}$\\  According to
the meaning of Higgs field, this is the Lagrangian density for Higgs
field. Thus we obtain the Lagrangian density for Higgs field. let $k=c$, then obtain:\\
$-|\nabla_{4}\rho _{vr}|^{2}=(\partial_{u}\rho_{vr})^{\dagger}\partial^{u}\rho_{vr}$,
$u=1, 2, 3, 4$\\
where, $\partial_{u}$, $\partial^{u}$ are the covariant form derivative and the contravariant 
form derivative, thus we obtain:\\
\begin{equation}L_{v}=W_{zcv}-\beta_{0}(\partial_{u}\rho_{vr})^{+}\partial^{u}\rho_{vr}
+\beta_{1}|\rho_{vr}|^{2}+\beta_{2}|\rho_{vr}|^{4}\end{equation}\label{20}\\
this is the form of the Lagrangian density of Higgs field which is used in the electroweak 
unified theory(the difference between them is a constant only). Though their forms are 
similar, the component numbers in the isospace are different(In the electroweak unified 
theory, how to obtain the Lagrangian density of Higgs field has not been
explained, so is it in supersymmetric theories). In my follow-up 
papers, I will discuss the essence of time, the origin of mass, the
origin of spin, the quantization of  electric
charge;  the four basic interactions  and their unification
from the angle of the level of disorder.
\section{References}
$[1]$:  Nir Polonsky, Supersymmetry:Structure and Phenomena,
Extensions of Standard Model, Springer(2001).\\
\indent  Gordon Kane, Supersymmetry, Squarks, Photinos and Unveiling of
the Ultimate Laws of Nature. Perseus Publishing, Cambridge Massachusetts(2000)\\
$[2]$:  Amitabha Lahiri, Palash B. Pal, Quantum Field Theory, CRC Press (2000)\\  
$[3]$:  Edited and translated by Fan Dai Nian, Zhao Zhong Li, Xu Liang Ying.The 
Collected Papers of Einstein, Vol. 2. Published by the Commercial Press (1983).\\
$[4]$:  Original Author, Huang Kun, Revised by Han Qi, Solid StatePhysics, Senior Education 
Press (1985). \\
\indent  W.Greiner B.M$\hat{u}$ller J.Rafelski, Quantum Electrodynamics of Strong 
Fields, Springer-Verlag,Berlin Heidelberg New York okyo(1985).\\
\end{document}